\documentclass[aps,prl,preprint]{revtex4}
\usepackage{CJK}
\usepackage{color}
\usepackage{amssymb}
\usepackage{graphicx}

\begin{document}

%\section{Table of contents}
%
%\includegraphics[scale=1.0]{figures/TOC}
%
%We measure the contrast of individually diffusing Au nanospheres in a tunable nanofluidic slit. Using an \textit{in-situ} calibration of the contrast we obtain a gap distance dependent measurement of the particle height probability distribution $p(h)$.
%
%\newpage

%%%%%%%%%%%%%%%%%% title page information %%%%%%%%%%%%%%%%%%
\begin{CJK*}{GB}{}
\title{\textit{In-situ} contrast calibration to determine the height of individual diffusing nanoparticles in a tunable confinement}

\author{Stefan Fringes$^1$, Michael Skaug$^1$ and Armin W. Knoll$^{1}$}

\email{ark@zurich.ibm.com}

\affiliation{$^1$IBM Research - Zurich, S\"{a}umerstr. 4, CH-8803 R\"{u}schlikon, Switzerland}

\begin{abstract}
	We study the behavior of charged spherical Au nanoparticles in a  nanofluidic slit as a function of the separation of the symmetrically charged confining surfaces. A dedicated setup called the nano-fluidic confinement apparatus (NCA) allows us to parallelize the two confining surfaces and to continuously approach them down to direct contact. Interferometric scattering (iSCAT) detection is used to measure the particle contrast with 2 ms temporal resolution. We obtain the confinement gap distance from the interference signal of the glass and the oxide-covered silicon wafer surface with nanometer accuracy. We present a three parameter model that describes the optical signal of the particles as a function of particle height and gap distance. The model is verified using nanoparticles immobilized at the glass and the substrate surface. For freely diffusing particles, the envelope of the particle signal as a function of gap distance and the known particle height at tight confinement is used to calibrate the particle signal \textit{in-situ} and obtain all free model parameters. Due to the periodic contrast variation for large gap distances we obtain a set of possible particle heights for a given contrast value. For a range of small gap distances this assignment is unique and the particle height can be measured directly with high accuracy. The high temporal resolution allows us to measure the height occupation probability which provides a direct link to the free-energy landscape the particles are probing via the Boltzmann distribution. Accordingly by fitting the results to a physical model based on the linear superposition approximation (LSA), the physical parameters governing the particle-glass interaction are quantified.
\end{abstract}

\maketitle


\end{CJK*}

%%%%%%%%%%%%%%%%%%%%%%%%%%  body  %%%%%%%%%%%%%%%%%%%%%%%%%%
\section{Introduction}
Confinement of particle motion plays a key role in many technological and natural systems.  In particle self-assembly for example, geometric confinement can be used to template desired structures or manipulate the available phase space or system entropy to tune interactions \cite{Grzybowski2009,Grzelczak2010,Anders2014}.  Likewise, the reduction of particle entropy imposed by confinement leads to novel entropy-controlled particle transport, which might serve as an efficient particle-separation mechanism \cite{Reguera2006,Burada2009}.  Fundamentally, confined particle motion is not as well understood as its bulk counterpart, despite its importance in nanofluidics \cite{Bocquet2014}, biological systems \cite{Baum2014}, porous materials \cite{Karger2013} and confined reactions \cite{Li2014}. To probe the fundamentals of confined particle transport, the ideal experiment would allow 3D particle tracking and precise control over the confinement conditions.

A variety of methods have been developed to track the 3D motion of nanoparticles or molecules.  Widefield methods, such as point-spread function engineering \cite{Pavani09apl,Shechtman15nanlet}, holographic \cite{Verpillat11optexp}, off-focus \cite{Speidel03}, astigmatic \cite{Kao1994,Huang2008} and multifocal-plane \cite{Juette08natmet,Ram08,Ram2012} microscopy, provide high-resolution 3D position detection.
%Active feedback confocal systems overcome some of these limitations, but still only report relative changes in z-position and can only track a single particle \cite{Han2012}.
Another approach is to use the interference of light reflected by the particles and a reference surface to determine the absolute particle-surface distance. For example, using a tilted back-reflecting mirror and dynamic illumination, the height of a fluorescence labeled static lipid bilayer was determined with a resolution of $\approx 3\,$nm  \cite{Kerssemakers14nanlet}. In a more general approach, interferometric scattering detection (iSCAT) was developed to measure the 3D positions of nano-objects with high spatio-temporal resolution \cite{Jacobsen06optexp,Kukura09natmet}. Because of the interference contrast the method scales very favorably for small reflective particles and/or for high temporal resolutions \cite{Piliarik14natcom}.
%Calibration of the optical signal is usually done by a measurement of the average optical response from particles immobilized at known heights in the system.
Combining iSCAT with nanofluidic confinement, nanosized objects were tracked in 3D and could be confined to geometric traps for hours \cite{Krishnan10nature, Gerspach15}. Trapping was achieved by shaping a topographical recess in one of the confining surfaces and exploiting the repulsive force between the like charged particle and surfaces \cite{Krishnan10nature}. Using a nanopipette in a scanning probe microscopy system it was shown that the trapping stiffness could be increased by increasing the confinement \cite{TaeKim14natcom}.
%In the nanofluidic slit, charged particles were stabilized by an electrostatic repulsive force from the confining surfaces due to the dissociation of ions and the formation of an electric double layer. A recess in one of the confining surfaces shaped the local potential minimum and thus overcame the randomizing effect of Brownian motion and trapped nanosized objects for several hours  \cite{Krishnan10nature}.
In these experiments the optical particle contrast intensity is typically calibrated by obtaining average intensity values of particles sticking to interfaces at known heights. The mean contrast of single trapped particles is then used to adapt this intensity scale to the individual particle level \cite{Mojarad13optexp}.
%The resulting 3D probability distribution of particle positions directly mapped the potential landscape experienced by the particles.

In this work, we calibrate the contrast of individual nanoparticles \textit{in-situ}. Our experimental NCA setup combines tunable confinement with iSCAT imaging. It allows us to track the three-dimensional position of single diffusing $60\,$nm gold nano-spheres with short illumination times. %The tunable slit height allowed us to measure particle behavior as a function of confinement and enabled in-situ calibration of each nanoparticle, which is important for a heterogeneous population of particles.
In the first part of the manuscript, we develop a three-parameter optical model which describes the iSCAT contrast obtained for nano-spheres confined in a nanofludic gap as a function of gap distance and particle height. In the second part, we verify the model using 380 particles immobilized at both confining surfaces using the known height as model input.
%The parameters and their variation were consistent for all measured particles.
In the third part, we describe how we obtain all three free model parameters \textit{in-situ }for individual diffusing nano-spheres from a single measurement of the particle contrast as a function of surface separation.
%We found that the particles were confined at the center of the nano-fluidic gap which is expected due to the repulsion from the surfaces and the symmetry of the system and corroborates our model.
Lastly, we use the measured height distributions as input to a physical model in order to determine the physical quantities describing the interaction of the spheres with the confining surfaces.

\section{Method}

\begin{figure}
\centering
\includegraphics[scale=1.05]{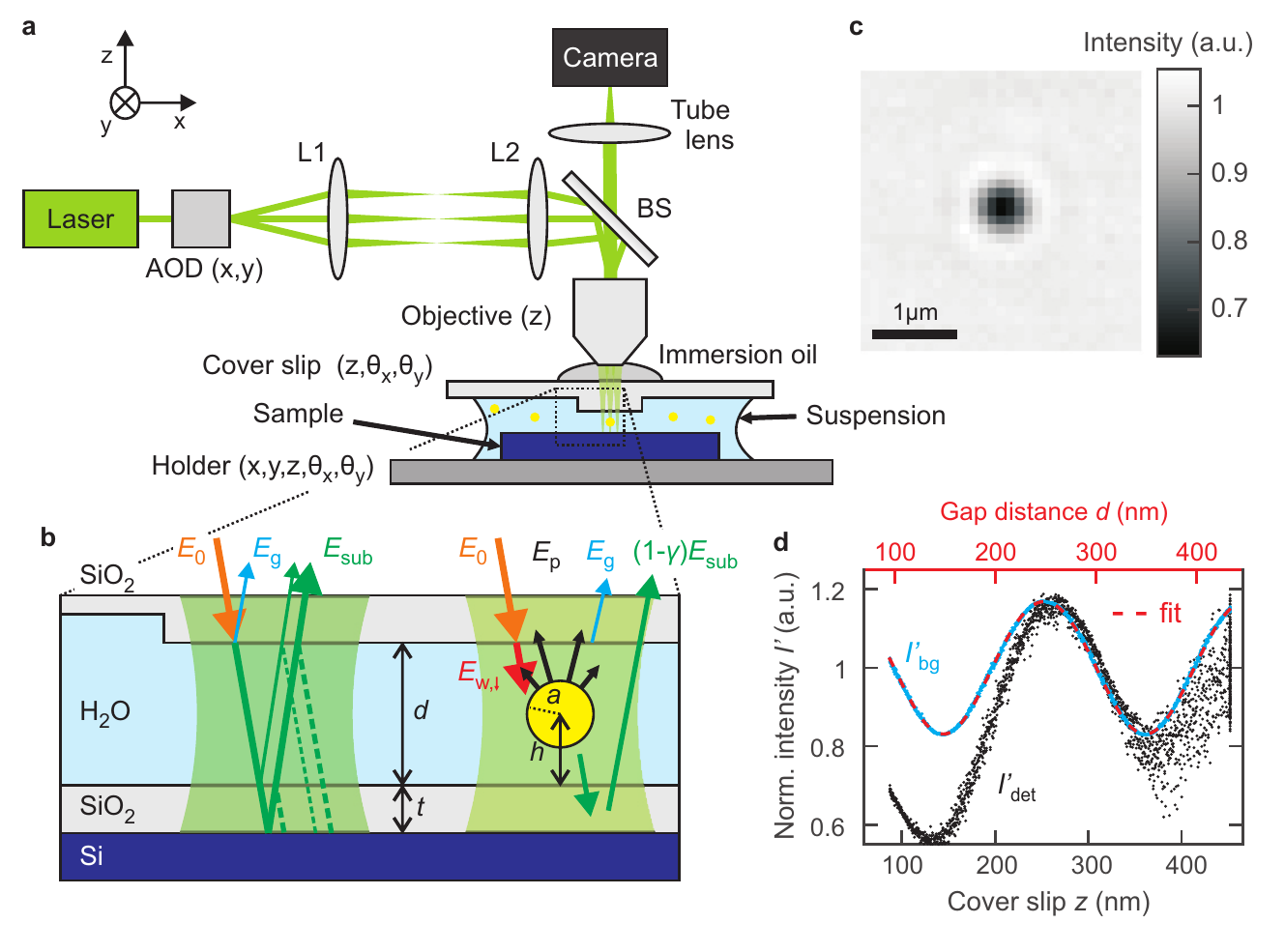}
\protect\caption{Experimental particle tracking setup implementing the nanofluidic confinement apparatus (NCA). \textbf{a}, The optical setup consist of a laser source, an acousto-optic deflector (AOD) and a telecentric system (L1 and L2) for scanned laser illumination of the sample through a beam splitter (BS) and a $\mathrm{NA}=1.4$, $\mathrm{M}=100x$ oil objective. The sample on the sample holder can be moved in 5 degrees of freedom ($x,y,z,\theta_{x},\theta_{y}$) and the cover-glass in 3 degrees ($z,\theta_{x},\theta_{y}$) relative to the objective. The gap between substrate and cover-glass is filled with nanoparticle suspension. A $\approx50\mu\mathrm{m}$ high mesa in the cover-glass ensures reproducible and obstacle-free motion of both surfaces down to intimate contact in the optically accessible region. b) Model used for signal interpretation without (left) and with a particle present (right). Without particle the incoming field $E_0$ is partially reflected at the glass-water interface ($E_g$) and partially transmitted into the gap, which it leaves after multiple reflections ($E_{sub}$). If a particle is present, it scatters and absorbs light, causing the background field to be reduced to ((1-$\gamma)\,E_{sub}$), see text for details. The particle reflection $E_p$ is modeled to be proportional to the field $E_{w, \downarrow}$ propagating towards the substrate in the water medium.
c) Optical image of a spherical Au nanoparticle with $a=30\,$nm radius. d) Typical signal recorded from the background and from a diffusing nanoparticle as a function of gap distance $d$. The dashed red line indicates a fit of the model to the background intensity. \label{fig:setup}}
\end{figure}

A schematic illustration of the NCA setup is shown in Fig. \ref{fig:setup}a. An aqueous suspension of $60\,$nm gold nanoparticles is confined within a nanofluidic slit formed by a glass coverslip and a silicon wafer piece substrate. The substrate is glued to a sample holder, which is positioned by a piezo-motor driven stage (Feinmess, KDT105) in the plane ($x,y$) of the table. In addition, three piezo motors (Picomotor, Newport) are used to adjust the tilt of the substrate with respect to the plane of motion. The cover-glass defines the top surface of the tunable nanofluidic gap. It is held by a metal frame that allows coarse tilt correction and positioning using magnets and screws. Linear piezo-stages (100$\,\mu$m, Nano-OP100, Mad City Labs) precisely position the cover-glass holder and the microscope objective with respect to the substrate. The cover-glass is patterned by optical lithography and hydrofluoric acid (HF) etching such that most of the surface facing the sample is recessed by $\approx 50\,\mu$m  except for a central mesa with a diameter of $\approx$ 100 $\mathrm{\mu m}$. The mesa provides good optical access to the fluidic slit and ensures that both surfaces can be approached continuously down to intimate contact.

Particle tracking and imaging are done using a typical iSCAT \cite{Jacobsen06optexp,Mojarad13optexp,Kukura09natmet} detection setup. A continuous-wave laser ($532.1\,\pm\,0.3\,$nm, Samba, Cobolt) with a beam diameter of $\approx 0.7\,$mm illuminates a selected sample area via raster scanning the focused laser beam by a two-axis acousto-optic deflector (AOD). Focusing and imaging are done using a 100x, 1.4 NA oil-immersion objective (Alpha Plan-Apochromat, Zeiss). A telecentric system is used to image the AOD (DTSXY, AA Opto-Electronic) deflection plane onto the back focal plane of the objective. Using a high frame rate camera (MV-D1024-160-CL-12, Photon Focus), the imaging rate is 500 frames per second (FPS) for imaging a $40 \times 40\,\mu$m$^2$ area using $400 \times 400$ pixels. During each image acquisition, the laser is raster scanned either once to minimize the illumination time of the particles or four times to optimize the background noise. Using one raster scan of the laser, the time a particle is illuminated can be estimated from the laser line spacing of $200\,$nm, an approximate spot size of the laser of $\lesssim 2\,\mu$m and the time per laser line of $10\,\mu$s to be $\sigma_{illu}\lesssim\,100\,\mu$s. Raster scanning the laser four times per frame results in an illumination time of $\sigma_{illu}\approx1.5\,$ms for a $\sigma=2\,$ms camera exposure \cite{Mojarad13optexp}.

Fig. \ref{fig:setup}b schematically illustrates the pillar-substrate nanofluidic region. As measured by ellipsometry, a $t = 51.0 \pm 0.5\,$nm thick layer of SiO$_2$ has been deposited by evaporation on both surfaces, which ensures that the same charge density is present on the confining surfaces. The axial position, $z$, and thus the gap distance, $d$, between cover-glass and sample is controlled with sub-nanometer precision. The average background intensity $I_{bg}$ recorded as a function of axial position $z$ is shown in Fig. \ref{fig:setup}d. The z-axis has been calibrated using interferometry, see Appendix A. The signal is dominated by the quasi-sinusoidal intensity variation originating from the interference between light reflected from the cover-glass and the substrate, see Appendix B. To determine the tilt of the two surfaces ($\Theta_{x}$, $\Theta_{y}$), we measure the interference signal separately for the four corners of the illuminated area. The phase shift of these intensity signals is directly related to the respective height difference and can be measured with nanometer accuracy. Using the three actuators in the sample holder, we achieve a planarisation of the two surfaces of better than $1\,$nm per 10$\,\mu$m.

An exemplary image of a $60\,$nm gold nanosphere (BBI Solutions) used in this study in a $90\,$nm wide nanofluidic gap is shown in Fig. \ref{fig:setup}c. The manufacturer describes the colloidal suspension as having an optical density of 100, a particle density of $\approx 2.6 \cdot 10^{12}$ particles per ml, a mean particle diameter of $59.8\,$nm with a size variation below 8\% and gold chloride concentration of 1\%. The particles are stabilized by the charged organic molecule citrate and a zeta potential of $-58\,$mV was measured for a 1:150 diluted suspension using a Malvern Zetasizer. For the experiments the suspension  was diluted 1:30 in deionized $\mathrm{H_{2}O}$ ($18\,\mathrm{M\Omega cm^{-1}}$) to reduce the ion concentration and obtain a particle density that was suitable for tracking. For the experiments using immobilized particles at the cover-glass, the particles were deposited by droplet drying. SEM inspection after the drying process revealed residues around the particles, which are likely due to the deposition of citrate upon water evaporation. Deionized $\mathrm{H_{2}O}$ was used during imaging.

In all experiments the position of the objective was optimized to ensure that all particles displayed maximum contrast for strongest confinement. The recorded frames were processed to remove the camera's dark intensity counts and to normalize the signal, see Appendix C. Typically, the height of the objective was fixed when imaging particles. When imaging particles immobilized at the cover-glass we observed that the image contrast was dependent on the relative position of the objective, i.e. if the objective was held at a fixed position or if it was moved in parallel with the cover-glass to keep the particles in focus. The effect and how it was corrected for the moving particles is described in Appendix D.

After Radial symmetry-based tracking was used to detect the center of the particles, a method that achieves similar accuracies as Gaussian fitting \cite{Parthasarathy12natmet}. Depending on the height of the particle, the central contrast of a particle can be either negative, positive or zero. We found that a particle with no contrast in the center can still be localized by the non-vanishing diffraction rings, which enabled tracking of the particle through the zero-contrast region (see Movie 1). Here the ability of the tracking algorithm in detecting the center of such arbitrary centro-symmetrical features is important. The center of the particle was localized with sub-pixel accuracy, and Gaussian fitting was used to determine the central particle intensity.

\section{Results}

\subsection{Optical Parametric Model}

The normalized measured intensity values for a single diffusing particle as a function of the axial piezo z-position controlling the separation of the two confining surfaces is depicted by the black scatter plot in Fig. \ref{fig:setup}d. In addition, the signal of the normalized background intensity (see Appendix C) is shown in blue. From the data, two critical aspects are immediately apparent: 1) The overall amplitude of the particle signal oscillations is roughly twice the amplitude of the background signal oscillations. 2) The particle intensity is less than the background intensity for most data points. These observations cannot be explained by the model often used for interpreting iSCAT intensity data in which the reflectivity of the glass surface is ignored \cite{Krishnan10nature,Mojarad13optexp}:
\begin{equation}
\label{iSCAT}
I_{det} = |E_{ref} + E_p|^2 = I_0 (|r_{ref}|^2 + |r_{p}|^2 + 2 |r_{ref} r_{p}| cos \phi),
\end{equation}
where $I_{det}$ is the detected intensity at the particle position and $\phi$ is the relative phase shift of the two interfering light fields from the reflective reference surface, $E_{ref}$, and the particle, $E_p$. Eq. (\ref{iSCAT}) predicts a particle intensity varying about a mean value $I_0 (|r_{p}|^2 + |r_{ref}|^2)$, which is greater and not less than the background intensity $I_0 |r_{ref}|^2$ and contradicts observation 2). Concerning observation 1), the background signal oscillations are of similar magnitude as the particle signal oscillations. This indicates that the fields originating from the glass and the particle reflections are of similar magnitude, and thus the glass reflection cannot be neglected.

Here we propose the following model as sketched in Fig. \ref{fig:setup}c. For the detected interference signal, we consider three interfaces as sources for light reflection, i.e., at the substrate $E_{sub}$, at the particle $E_p$ and at the glass surface $E_g$. That is we have two reference surfaces, the glass and the substrate. As the particle is located above the substrate, we have to consider the effect of the presence of a particle on the reflected substrate field, $E_{sub}$, because the light has to pass the particle before it arrives at the substrate and on the way back to the camera. To estimate the significance of this effect, we consider the effective areas involved when imaging the particle. The optical resolution of the microscope is given by the diffraction limit to be $\approx 0.61 \lambda/$NA$\,\approx 250\,$nm. As a particle image, we observe the interference signal of the light reflected from this area of the substrate with light scattered by the particle and the glass. Thus it is apparent that a $60\,$nm diameter Au particle scatters and absorbs a significant fraction of the relevant incoming light. This effect is even enhanced by the fact that the light interacts with a metallic particle close to its plasma frequency. We account for the reduced light intensity by introducing a parameter $\gamma$ describing the fraction of the incident field which is scattered or absorbed by the particle. As a consequence, the field emerging from the substrate at the particle location is reduced to $E_{sub,p} = (1-\gamma) E_{sub}$. The fraction $\gamma$ of the light interacting with the particle is partially scattered by the particle, transmitted to and collected by the objective. We term this light $E_p\,=\,\gamma E_p'$, where we rescale the unknown magnitude of the particle reflection with $\gamma$. Then we can write:
\begin{eqnarray}
  I_{bg} &=& |E_{sub} + E_g|^2 \nonumber \\
  &=& E_0^2 |r_{bg}|^2 = I_0 |r_{bg}|^2 \\
  I_{det} &\approx& |(1-\gamma)\,E_{sub} + E_p + E_g|^2 \nonumber\\
  &=& |(1-\gamma)\,(E_{sub} + E_g) + \gamma (E_p' + E_g)|^2 \label{equ3}\\
  &=& I_0 |(1-\gamma) r_{bg} + \gamma r_{p,g}|^2 \label{equ5},
\end{eqnarray}
where $r_{bg}$ is the reflection coefficient of the background far from a particle. It describes the multiple reflections of the light in the water medium, the transmission through the glass interface, and the interference with the light reflected by the glass-water interface. Because of the interference, $r_{bg}$ is now a function of the gap distance $d$. Also, $r_{p,g}\,=\,(E_p' + E_g)/E_0$ can be viewed as the reflection coefficient of the glass-particle system if only the light fraction $\gamma$ interacting with the particle is considered. It describes the light interacting with the particle and interfering with light reflected by the glass surface.

%Eq. \ref{equ5} describes the expected signal to originate from a fraction $1-\gamma$ of the background field and its interference with a fraction $\gamma$ of the particle-glass field.
%The normalized particle contrast $I_{det} - I_{bg}$ now depends on the gap distance $d$. The contrast amounts to $\left((1-\gamma)^2 \langle|r_{bg}|\rangle^2 + \gamma^2 \langle|r_{p,g}|\rangle^2\right)/\langle|r_{bg}|\rangle^2\,=\,(1-\gamma)^2 + \gamma^2 \langle|r_{p,g}|\rangle^2/\langle|r_{bg}|\rangle^2$, which may assume values smaller than one as observed in the experiment.

We model the background reflection by employing a transfer-matrix method \cite{Bornbook, SerneliusBook}. This method is used to determine the reflectivity of plan-parallel multi-layer stratified media stacks. Each layer of the stack is characterized by a reflection and a propagation part for the light entering the layer. For light entering layer 2 from layer 1 and propagating to the next interface to layer 3, the transfer-matrix is calculated by
\begin{equation}
M_{12}=\frac{1}{t_{12}\left(\theta_2\right)}\left[\begin{array}{cc}
1 & r_{12}\left(\theta_2\right)\\
r_{12}\left(\theta_2\right) & 1
\end{array}\right] \left[\begin{array}{cc}
e^{i k_2\left(\theta_2\right) d} & 0\\
0 & e^{-i k_2\left(\theta_2\right) d}
\end{array}\right],
\end{equation}
where $t_{12}\left(\theta_2\right)$ and $r_{12}\left(\theta_2\right)$ are the transmission and reflection coefficients, while $k_2\left(\theta_2\right) = 2\pi n_2 \cos\left(\theta_2\right)/ \lambda $ is the z-component of wave vector at incident angle $\theta_2$ in medium 2, and $d$ is the thickness of layer 2. The total matrix for the background signal is composed of the matrices for the glass-water interface, $M_{ow}$, the water-oxide interface, $M_{wo}$, and the oxide-silicon interface, $M_{os}$. For the latter it is assumed that all the light not reflected from the interface is absorbed in the bulk of the silicon wafer, i.e., the propagation part of the transfer matrix is omitted.
The total transfer matrix describing the entire stack is then given by $M_{bg}=M_{ow}M_{wo}M_{os}$ and the reflection coefficient of the stack is obtained from
\begin{equation}
\label{rbg}
r_{bg}(d)\,=\,M_{bg}^{2,1}\,/\,M_{bg}^{1,1},
\end{equation}
where the indices denote the respective matrix elements of $M_{bg}$. The normalized background intensity is then given by
\begin{equation}
\label{Ibg}
I'_{bg}(d) = \frac{I_{bg}(d)}{\langle I_{bg} \rangle} = \frac{|r_{bg}(d)|^2}{\langle |r_{bg}|^2 \rangle},
\end{equation}
where the averaging is done over one period of the interference signal, i.e., a distance range of $200\,$nm.

In order to calculate the gap distance from the measured normalized background intensity we first calibrate the piezo-stage in air (see Appendix A). Second, we have to consider that the focusing of the laser beam leads to light rays with finite incident angles. Because of the strong underfilling of the objective we approximate the illumination by calculating an effective incident angle $\theta_{eff} = 5.9 ^\circ$ (see Appendix B) obtained from a measurement of the interference signal in air. We use Snell's law to propagate $\theta_{eff}$ into the dielectric layers and Fresnel's formulas to calculate the reflection and transmission coefficients. The refractive index of $n_{H_2O} = 1.33$ for water and $n_{Si} = 4.14$ for silicon are obtained from literature, while the refractive index of silicon oxide, $n_{ox}\,=\,1.476\,\pm\,0.006$, and its thickness, $t = 51.0 \pm 0.5\,$nm, see methods.

The dashed red line in Fig. \ref{fig:setup}d is a fit of Eq. \ref{Ibg} to the data using two free parameters: a linear scaling factor c, and a relative offset $z_{offs}$ for the conversion of the z-axis to gap-distance values, $d = cz + z_{offs}$. The scaling factor c is required because of pressure effects from the confined droplet between compliant cover-glass and sample. We obtain c = $0.961$, i.e., a deviation of about $4\%$ from a 1:1 correspondence. For the results discussed in the following sections, we use the background intensity signal for a calibrated measurement of the gap distance.
%That is we determine the gap distance on a pixel by pixel basis by fitting equation \ref{Ibg} for every pixel using only the gap distance offset as fit parameter. The result is an image of the optical gap distance $d(x,y)$, see Fig. Sx.

To complete our model, we still need to approximate the reflection coefficient for the glass-particle system, $r_{p,g}$. For this we need to compute the light field scattered by the particle. We assume that the amplitude of the scattered field is proportional to the amplitude of the incoming field within the nanofluidic gap, i.e. the wave propagating in the water medium towards the substrate, $E_{w,\downarrow}$, see Fig. \ref{fig:setup}c. The complex amplitudes of $E_{w,\downarrow}$ (and $E_{w,\uparrow}$) at the glass-water interface are determined from the transfer-matrix formalism as follows:
\begin{equation}
\label{Epropagatingdown}
\left[\begin{array}{cc}
E_{w,\downarrow}\left(\theta_w\right)\\
E_{w,\uparrow}\left(\theta_w\right)
\end{array}\right] = \left(\frac{1}{t_{g,w}\left(\theta_w\right)}\left[\begin{array}{cc}
1 & r_{g,w}\left(\theta_w\right)\\
r_{g,w}\left(\theta_w\right) & 1
\end{array}\right]\right)^{-1}
\left[\begin{array}{cc}
1\\
r_{bg}\left(\theta_w\right)
\end{array}\right] E_0.
\end{equation}
The incoming field $E_{w,\downarrow}$ propagates a distance $(d-h)/\cos\left(\theta_w\right)$ towards the particle (centered at height $h$), is reflected according to a complex reflection coefficient $r_p$, propagates back to the glass interface, and is transmitted through the water-glass interface. We neglect any light transmitted through the particle and reflected by the substrate or forward-scattered light from the upward propagating wave $E_{w,\uparrow}$. The back-scattered light is emitted in several directions, transmitted through the glass interface, and arrives at the camera pixel, where it interferes with itself and with the light reflected from the glass surface. We incorporate all of the losses and phase shifts inherent in the path towards the camera in the complex scattering coefficient $r_p$ and thus obtain the following relation for the reflection coefficient of the particle-glass system:
\begin{equation}
\label{rpg}
r_{p,g} \approx r_g + E_{w,\downarrow}/E_0 e^{2 i k_w\left[d-h\right]} p e^{i \phi_0},
\end{equation}
where we assign an amplitude $p$ and an accumulated scattering phase $\phi_0$ to the particle reflection coefficient $r_p$, and $k_w$ is the z-component of the wave vector incitent at an angle $\theta_w$ in water.
%As stated above, $r_{p,g}$ only considers the light interacting with the particle. That is, we expect the reflected amplitude $p$ to be smaller than unity because some light will be absorbed and not all the scattered light will be collected by the objective.

Equations (\ref{equ5}), (\ref{rbg}), and (\ref{rpg}) describe the optical iSCAT signal of a particle in a nanofluidic gap. The excellent fit of the background intensity to the model for $I_{bg}$ allow us to retrieve all the values for the transmission and reflection coefficients concerning the glass and the substrate interference. In order to determine the particle height $h$ [Eq. (\ref{rpg})] the remaining unknown parameters for describing the particle signal have to be specified, the fraction $\gamma$ of the incoming light interacting with the particle [Eq. (\ref{equ5})], and the particle's scattering amplitude $p$ and the effective scattering phase $\phi_0$ [Eq. (\ref{rpg})].

\subsection{Model verification using immobilized particles}

\begin{figure}
\centering
\includegraphics[scale=1.4]{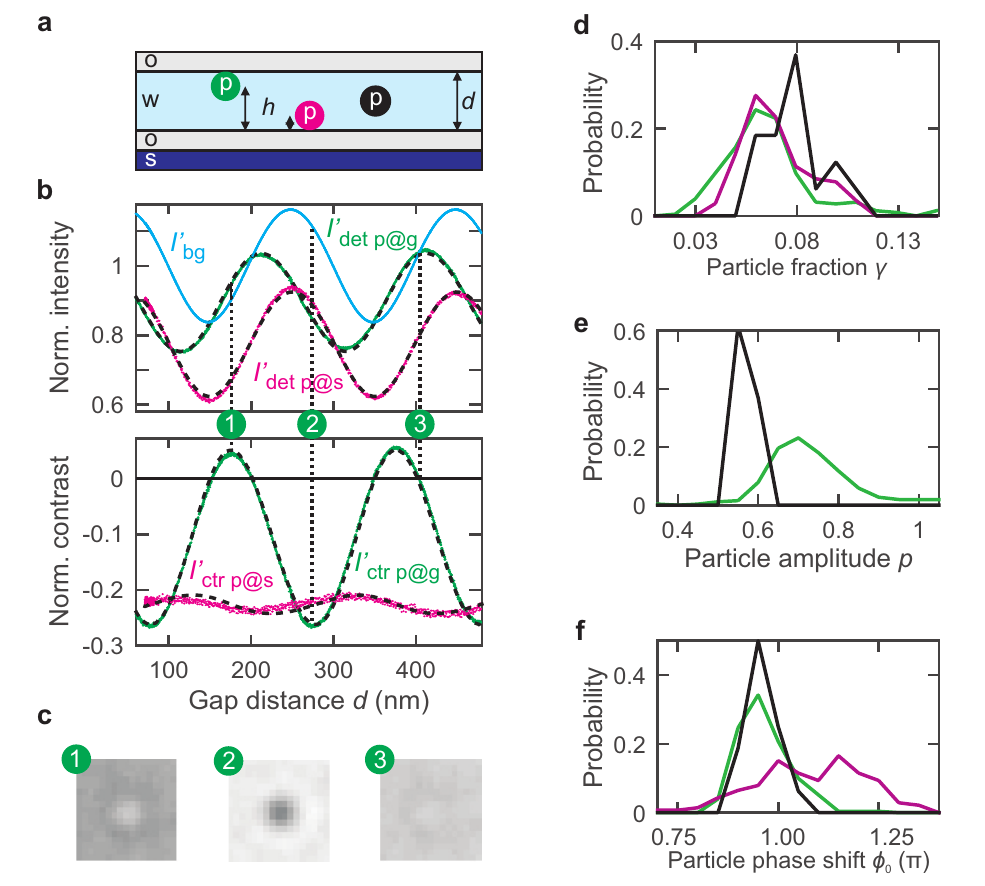}
\protect\caption{Verification of the optical model using particles fixed to the cover-glass and substrate. \textbf{a)} Schematic of particles fixed to the cover-glass (green), to the substrate (magenta) or freely diffusing (black). \textbf{b)} Examples for measured intensities (top) and contrast signals (bottom) for background and particle intensities for particles attached to the cover-glass (green) and the substrate (magenta). Dashed black lines represent fits according to the optical model. \textbf{c)} Sample images for maximal positive, maximal negative and zero contrast. Tracking was possible through zero contrast because of the nonzero radial signal at zero center contrast. \textbf{d-f)} Statistics for the optical parameters interaction fraction $\gamma$, intensity factor $p$ and phase $\phi_0$ describing the signal of all particles measured. \label{fig:model} }
\end{figure}

We tested the model derived in the preceding section by a measurement of $I_{det}$ and $I_{bg}$ for our particles of radius $a\,=\,30\,$nm immobilized at the surface of the cover-glass and at the substrate-water interface (Fig. \ref{fig:model}a), thus fixing the particle height, $h$, to $h = d-a$ and $h = a$, respectively. A total of 380 particle signals were analyzed, 241 attached at the cover-glass and 139 at the substrate. The normalized raw particle signal $I'_{det} = I_{det}/\langle I_{bg}\rangle$ and the normalized contrast $I'_{ctr} = (I_{det}-I_{bg})/\langle I_{bg}\rangle$ of the particles are shown in Fig. \ref{fig:model}b for two selected particles, one sticking to the substrate (magenta) and the other sticking to the glass (green). In addition, $I'_{bg}$ measured in the neighborhood of the particles is displayed in blue in Fig. \ref{fig:model}b.

We note that the signal for the particles fixed to the cover-glass was obtained by moving the objective in parallel with the cover-glass in order to keep the particles in focus, see also methods. If we measured the signal with a fixed position of the objective we observed an additional increasing phase with displacement of the particle out of focus, likely due to an accumulated Gouy phase. The effect is described in Appendix D and was compensated for the measurement of the freely diffusing particles.

The dashed lines in Fig. \ref{fig:model}b correspond to the best fit to the data according to our model. For particles fixed to the cover-glass, all three parameters were obtained from the fit. For particles sticking to the substrate, the raw signal is dominated by the reduced background signal $(1-\gamma)^2 I_{bg}$, evidenced by the vanishing relative phase of $I_{bg}$ and $I_{det}$. %Consequently the contrast signal is shifted by approximately 180 degrees.
The particle signal adds to the substrate signal with a fixed phase relation because of the fixed distance between particle and substrate. Therefore for particles fixed to the substrate, the phase $\phi_0$ and amplitude $p$ could not be obtained simultaneously. Here we fixed $p$ to an average value of $p=0.65$ as obtained from the other measurements and we used the particle scattering phase $\phi_0$ as fit parameter.

The histograms shown in Fig. \ref{fig:model}d-f represent the statistical distributions of the free parameters $\gamma$, $p$, and $\phi_S$, respectively, as obtained from the fit to a total of 380 particles. Histograms for particles immobilized on the cover-glass, on the substrate and freely diffusing (16 particles, evaluation see next section) are shown in green, magenta and black, respectively. We obtained histograms of similar width for all measurements. It is apparent that the values for the freely diffusing particles, in particular for $p$, are slightly shifted in comparison to those obtained obtained for the fixed particles. We attribute these deviations to the fact that during the drying process the interface between the particle and the cover-glass was filled with organic residues (see methods) which have a higher refractive index than water, resulting in higher reflectivities. As mentioned above, for the particles fixed to the substrate, we had to fix $p$. Because the phase now has to adapt for the total signal change, the values obtained for $\phi_0$ have a much wider distribution than those obtained for particles fixed at the glass surface or the freely diffusing particles.

In summary, fits to all 380 particle signals could be obtained with high fidelity. The standard deviation of the fits to the intensity data $I_{det}$ (see Fig. \ref{fig:model}b) for the particles sticking to the cover-glass surface was $(7.9 \pm 2.9) \,10^{-3}$ and for the particles sticking to the sample $(1.3 \pm 0.4) \,10^{-2}$ of the normalized intensity, which quantifies the good agreement of the model to the data. In addition, the values obtained for fixed and moving particles are in good agreement for all particle sets investigated. We interpret the excellent agreement as a first indication of the validity of our model.

\subsection{\textit{In-situ} contrast calibration for freely diffusing nano-spheres}

\begin{figure}
\centering
\includegraphics[scale=1.2]{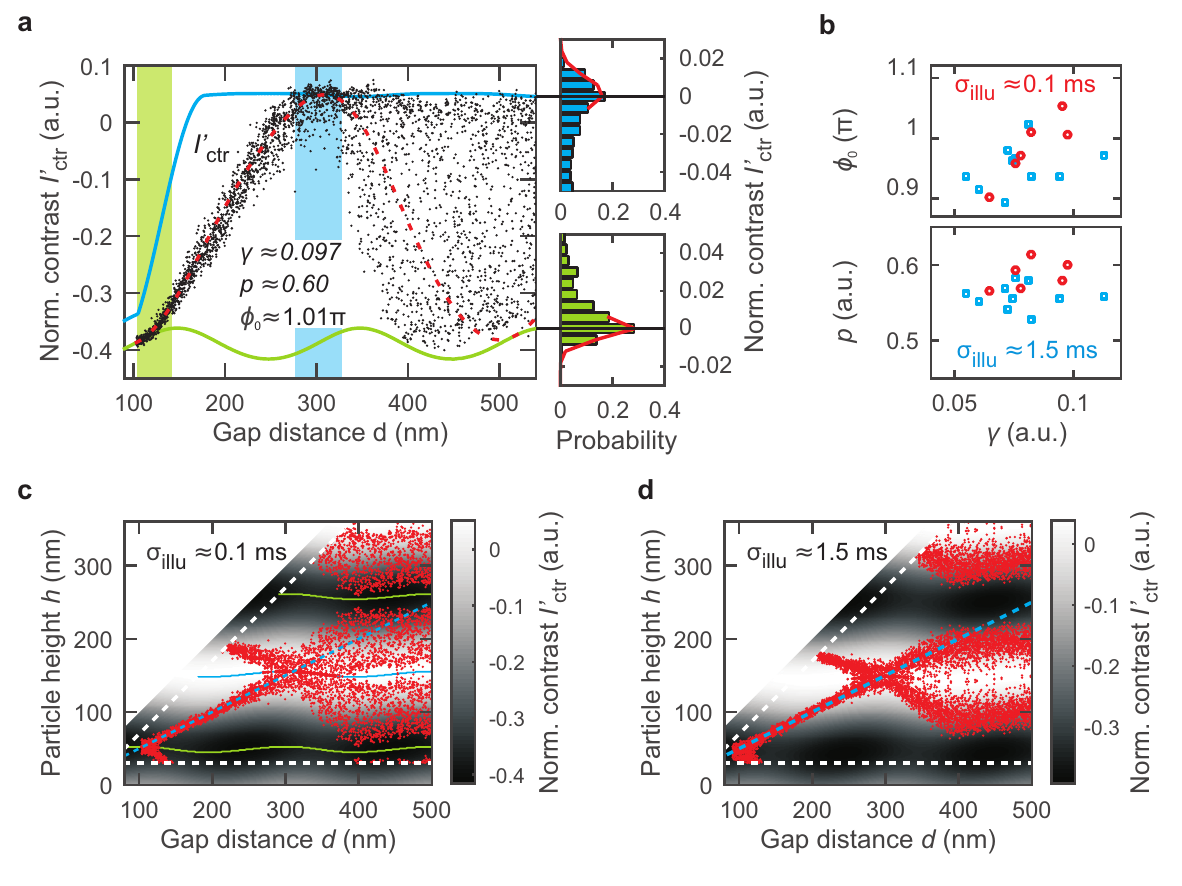}
\protect\caption{Measurement of the particle height for individual suspended nanoparticles as a function of gap distance. \textbf{a)} Measured contrast signal (black) of a single diffusing particle and envelopes of the simulated maximum (blue) and minimum (green) particle contrast for varying gap distance $d$. The dashed red line illustrates the simulated contrast for a particle at mid-gap position. The histograms are calculated for the marked regions (green and blue) and show the frequency of observing a particle contrast relative to the contrast value given by the envelope. Owing to noise, contrast values outside the envelope are observed, which are characterized by Gaussian distributions (red lines) having widths of $\sigma = 0.0085$ and $\sigma = 0.0055$ for the top and the bottom envelope, respectively. \textbf{b)} Correlations of the particle phase $\phi_0$ and the effective scattering amplitude $p$ with the fraction $\gamma$ for illumination times of $\approx 100\,\mu$s (red circles) and $1.5\,$ms (blue squares). \textbf{c)} and \textbf{d)} Attributed particle heights obtained for illumination times of $\approx 100\,\mu$s in c) and $1.5\,$ms in d). The gray-scale images visualize the simulated contrast for varying gap distance $d$ and particle height $h$. The blue and green lines denote the position of maximum and minimum simulated contrast, respectively. The red data points denote the height values obtained from the corresponding simulated contrast. The height values are restricted to the physically possible values within the gap given by the finite particle radius (dashed white lines). The dashed blue line illustrates the position of the gap center.	 \label{fig:height}}
\end{figure}

The measured contrast values $I'_{ctr}$ of a freely diffusing particle confined in the gap, recorded at varying gap distance, $d$, are depicted as black scatter plot in Fig. \ref{fig:height}a. The data were recorded at 500 FPS and an illumination time of $\approx 100\,\mu$s, see methods. The normalized contrast values are contained in a well-defined range $-0.4 \leq I'_{ctr} \leq0.05$, which we will use to determine the three unknown parameters $\gamma$, $p$, and $\phi_0$ on an individual particle level.

%a frame rate of 500 FPS and with a illumination time per frame of $\approx 100\,\mu$s.
%Up to gap distances below $\approx300\,$nm the signal has a narrow distribution and broadens for larger gap distances.

Using the optical model from Eqs. (\ref{equ5}), (\ref{rbg}), and (\ref{rpg}), and using a set of trial parameters $\gamma$, $p$, and $\phi_0$, we calculate the range of expected particle contrasts for all possible heights within the gap, $a\leq h \leq d-a$.  The minimum and maximum of the calculated contrasts define a theoretical envelope for possible particle contrasts observed experimentally.  These extremal contrasts are depicted in Fig. \ref{fig:height}a as green and blue lines.  The fitting procedure involves optimizing the three parameters,  $\gamma$, $p$, and $\phi_0$ such that the calculated envelope matches the range of measured contrasts in the experimental data. To characterize the deviation of the calculated envelope from the experimental data, we determined the contrast difference of the experimental data to the calculated envelope in a $50\,$nm wide band (blue and green areas in Fig. \ref{fig:height}a). The plotted histograms of the resulting values are shown on the right-hand side of Fig. \ref{fig:height}a. The center position of the blue band was obtained from the position of maximal contrast after smoothing of the contrast data.

Optimal values of $\gamma$, $p$, and $\phi_0$ were found by iteratively calculating the contrast envelope and histograms of the deviations until the peaks of the histograms were at zero. The optimized histograms and best fit envelope are shown in Fig. \ref{fig:height}a. This procedure was then repeated for each particle.

For an ideal noise-free measurement, the contrast histograms would comprise a sharp maximum feature given by the contrast envelope. In the measurements, we find a Gaussian tail, indicated by the red lines in the histograms in Fig. \ref{fig:height}a. We attribute this broadening to both noise from laser and camera and the fact that the particles may exhibit a time-dependent scattering cross-section due to shape asymmetry. The width of the Gaussian tails was typically less than 0.01 normalized intensity, which corresponds to errors of less than $2\,$nm for intermediate contrast values. The error in optimizing the fit may be approximated by the bin width of the histograms, which here was 0.01 normalized intensity. This corresponds to an additional systematic height error of approximately $2\,$nm for intermediate contrast values.

We found that the contrast envelope was predominantly related to $p$ and $\gamma$. The phase $\phi_0$ corresponds to an apparent change in particle height and had little effect on the envelope for large $d$. However for small $d$, the possible values for  $\phi_0$ are constrained to a narrow range because of the increasing confinement. Here we increased the confinement to roughly 90 nm, resulting in average distance of 15 nm from the particle surface to the confining interfaces. We prepared both surfaces with the same material, therefore we enforce the confinement condition further, assuming that the particle is located at the center of the gap, $h_c = d/2$ (see dashed red line in Fig. \ref{fig:height}a) for strong confinement, e.g. within the green band shown in Fig. \ref{fig:height}a.
%In the case of asymmetric surfaces one would likely observe a contact of the particle to one of the surfaces which could be used to determine the absolute phase $\phi_0$.

%Using the best fit $\gamma$, $p$, and $\phi_0$, we calculated the expected contrast for a particle in the gap center $h = d/2$ (red dashed line in Fig. \ref{fig:height}a). The line reaches a maximum, where also the number of data points concentrates at the upper boundary of the envelope. [In order to determine the corresponding gap distance we performed a moving average filter to the data and searched for the maximum value]?msk I thought the gap distance was measured independently using the background signal?.

%of measurement values to optimize the position of the upper envelope.
%The maximum of the filtered data curve is taken as the center gap distance for optimization. The width was adjusted to $50,\$nm, see blue shaded area in Fig. \ref{fig:height}a.

From the procedure described above, we obtain the parameters describing the optical response of the system for individual diffusing particles. This allows us to compare the parameters obtained for all individually diffusing particles. Fig. \ref{fig:height}b shows the correlations of $\phi_0$ and $p$ with $\gamma$. The phase $\phi_0$ may be interpreted as the distance between the particle origin and the actual plane of light reflection. Because the particle radius $a$ and the skin depth of gold are of similar magnitude, an average phase of about $\pi$ is observed. For a smaller particle (smaller value of $\gamma$), the reference plane is shifted towards the substrate interface and therefore the particle exhibits a more negative phase. The reflection amplitude $p$ changes very little with the particle size, see Fig. \ref{fig:height}b, lower panel. This is expected because $p$ describes the reflectivity of the light fraction that interacts with the particle. The minor increase of $p$ with $\gamma$ may be explained by the fact that for larger particles the scattering to extinction ratio increases \cite{Jain06}.

Once the optimal fit parameters have been obtained, the model can be used to calculate contrast values for given height values. The background of Figs. \ref{fig:height}c and d shows this calculated signal as a function of gap distance using the fit parameters obtained for the respective particles. The $200\,$nm periodicity of the contrast with the particle height is due to the $400\,$nm wavelength of light in water. The additional signal from the glass interface modulates the contrast along the axis of the gap distance, $d$, and also modulates the height values of extremal contrast by about $10\,$nm, see blue and green lines in Fig. \ref{fig:height}c. To obtain the height values for a measured contrast, we interpolated the calculated signal between the extremal values. Depending on the gap distance, this leads to single-valued or multi-valued results for the particle height. Contrast values measured outside the envelope function were set to the height values corresponding to extremal contrast.

The red data points in Fig. \ref{fig:height}c and d display the height values obtained that correspond to the measured contrast data recorded with an illumination time of $\sigma_{illu}\approx 100 \mu$s and $\sigma_{illu}\approx 1.5\,$ms, respectively. Because of the periodicity of the intensity along the $h$-axis, often more than one solution exists for possible particle heights, e.g., for gap distances $d \geq 220\,$nm. Therefore multiple "mirror" branches appear, and it is not possible to determine a priori which solution branch is the physical one. Note that we plotted data points only if the heights are within the range of accessible heights for the particle given by the gap distance, i.e. $a \leq h \leq d-a$, see dashed white lines in Fig. \ref{fig:height}c and d. Consequently for gap distances of $120\,$nm$ \lesssim d \lesssim 220\,$nm, the height data is single-valued and the corresponding branch can be interpreted as the physical branch. For this branch the heights are confined to a small band around the gap center denoted by the dashed blue line. In our system of a particle confined by symmetrically charged surfaces, this is precisely what is expected. Indeed, in the range $150\,$nm$\,<\,d\,<\,250\,$nm, we determine an average standard deviation of the average measured particle height from the gap center of $2\,$nm for all 16 particles measured,  which corroborates the validity of our model.

%At the end of this initial highly confined region at $d > 230\,$nm the distribution of measured particle heights widens and approaches the first maximum of particle contrast. For even larger
At gap distances of $d > 330\,$nm, the two measurements are distinctively different, with more confined bands of particle positions for the longer illumination time experiment shown in Fig. \ref{fig:height}d. At these gap distances, which are large compared to the Debye length of the system, the repulsive force from the surfaces is expected to become small (see also the next section). Therefore we expect a uniform distribution of possible height values along the h-axis around the gap center. This clearly is not the case in Fig. \ref{fig:height}d, where gaps with a width of $\approx 75\,$nm develop around the extremal contrast positions. The long illumination time effectively averages the apparent particle contrast for the fast and practically unconfined moving particles. The maximum contrast for a particle with extended illumination time corresponds to an average contrast value as the particle diffuses through the extremal contrast region. A simple calculation yields that the contrast gap of $\approx 75\,$nm observed in the experiment corresponds to a diffusion length of roughly $140\,$nm or a normal diffusion coefficient of $D_{\perp} \approx 6\,\mu$m$^2$/s. This is consistent with a diffusion coefficient $D=(k_B T )/(6\pi\eta a) \approx 7.2 \mu m^2/s$ and a diffusion length of $\Delta z = \sqrt{(2D\Delta t)} \approx 150\,$nm for an integration time (illumination time) of $\Delta t = \sigma_{illu} = 1.5$ ms as obtained for a freely moving particle using the Stokes-Einstein-equation \cite{Einstein1905}, and keeping in mind that the confinement slows down the diffusion \cite{Brenner61}. Similarly from the contrast gaps of $\approx 25\,$nm visible in Fig. \ref{fig:height}c, one can infer a diffusion length of $40\,$nm or an illumination time of 150$\,\mu$s, consistent with the experiment.

Clearly, a fast illumination time is required to obtain a faithful distribution of particle heights, in particular at the extremal values of the particle contrast. On the other hand, for freely moving particles and short illumination times the measured heights for intermediate contrast values have high accuracy because here the contrast scales linearly with height. In the next section, we will use this fact to characterize the physical properties of the system.

% $D_{\perp}/D = 0.78 $
% correction calculated by Brenner \cite{Brenner61,Kihm04}: 0.8542

%For the measurement employing the much smaller illumination time shown in Fig. \ref{fig:height}d, the gap is reduced to $\pm 12\,$nm, which corresponds to a diffusion length of just $40\,$nm or an illumination time of 110$\,\mu$s.

%For a freely diffusing Au nanoparticle of $2a = 60\,$nm diameter the diffusion coefficient $D=(k_B T )/(6\pi\eta a) \approx 7.2 \mu m^2/s$ and a diffusion length of $\Delta z = \sqrt{(2D\Delta t)} \approx 170\,$nm for a time step $\Delta t = 2$ ms can be calculated with the Stockes-Einstein-equation \cite{Einstein1905}.
%An explanation for the reduced tangential compared to the free diffusion coefficient is the increase of frictional resistance close to the presence of a solid wall. \cite{Brenner61}

%
%Experiments investigating the effect of defocusing on the phase shift have shown that it linearly increases with displacement of the particle out of focus $\Delta\phi=(h-z_{focus})\frac{11.7n_{w}}{\lambda}$ (Supplementary Information). This effect has been corrected the for the moving particles.

\section*{Application to study the particle surface interaction}
In this section, we use the measured height distributions to obtain the physical parameters describing the interaction of the charged sphere confined between the like charged surfaces. The forces on the particles in the nano-fluidic gap originate from the interaction of the electrostatic double-layers existing around the particle and close to the surfaces. To model the interactions, we use the linear superposition approximation (LSA) \cite{bell1970approximate, adamczyk1996role, Behrens2001} assuming a constant potential picture for a 1:1 electrolyte. The LSA approximation is strictly valid only for separations $x$ larger than the Debye length $\kappa^{-1}$ of the system; however, it has been shown that the errors at smaller separations are reasonably small \cite{adamczyk1996role}. For $x > \kappa^{-1}$, the electrostatic potential assumes small values ($\Phi < k_B T/q_e$) and thus can be described by a linearized Poisson-Boltzmann (PB) equation, $\psi(x) = \psi_{eff} \exp(- \kappa x)$. For the linear regime, superposition holds and a solution can be obtained for the overall system, in our case a particle and two surfaces, by adding the potentials obtained for the individual constituents. These individual potentials are calculated using the nonlinear PB equation. To match the nonlinear and linear solutions for $x > \kappa^{-1}$, the particle and surface potentials $\psi_P$ and $\psi_S$ are rescaled to so-called effective potentials, $\psi_{P,eff}$, and $\psi_{S,eff}$.
%Since the system is now described by linear equations superposition holds and the effective potentials of the isolated particles can simply be summed to obtain an approximate solution for the system.
Finally, the important parameters describing our system are the surface potentials of particle, $\psi_P$, and the surfaces, $\psi_S$, as well as the Debye length $\kappa^{-1}$.

\begin{figure}
\includegraphics{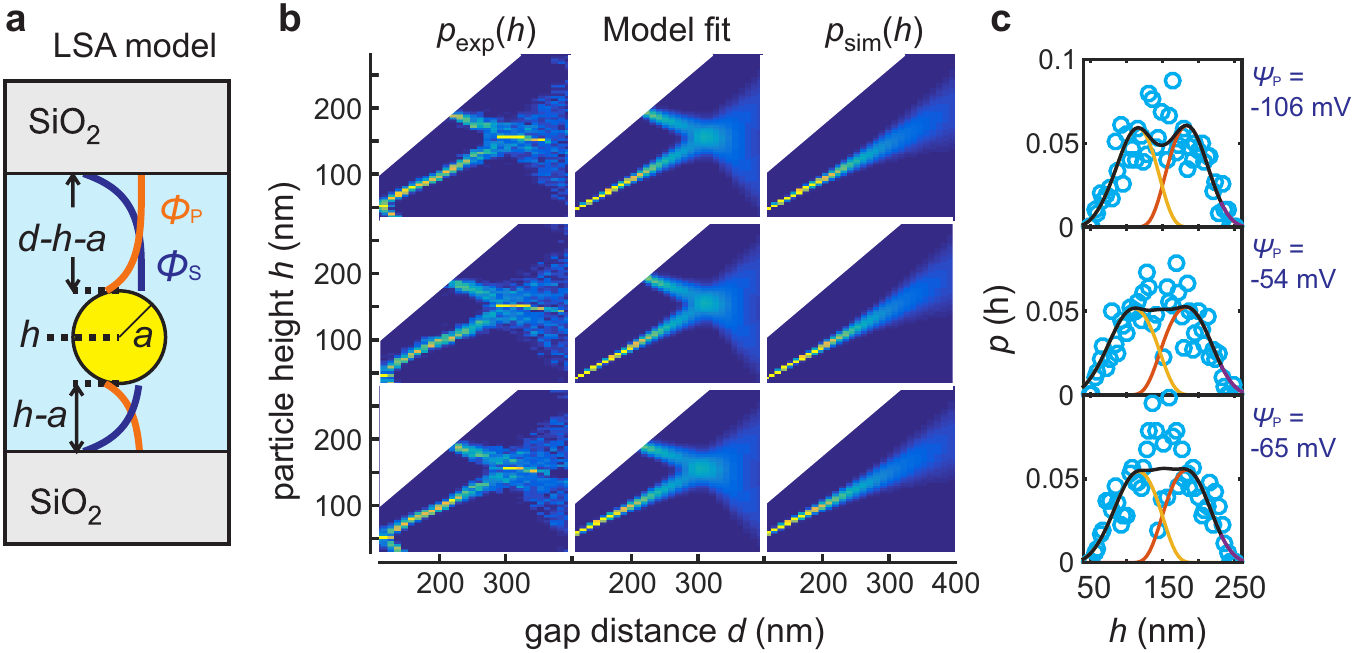}\protect\caption{Fit to a physical model based on the linear superposition approximation (LSA). \textbf{a} The model approximates the interaction potential by a superposition of particle potential $\Phi_{P}$ and surface potentials $\Phi_{S}$. The potentials are calculated for the effective separations between particle and SiO$_2$ surface of $d-h-a$ and $h-a$. \textbf{b} Comparison of experiment and simulation for three particles. Shown are experimentally obtained occupation probabilities $p_{exp}$ and model fit results including (Model fit) and excluding ($p_{sim}(h)$) the un-physical height positions for a given particle contrast. c) Measured histograms (blue circles) and fit results (lines) averaged in the range of gap distances of $350\, \mathrm{nm} < d < 400\,$nm. The simulated probability $p_{sim}$ is shown in red, the mirrored probability in yellow and the sum of both in black. \label{fig:LSA}}
\end{figure}

Fig. \ref{fig:LSA}a schematically depicts the particle in the nanofludic gap. We assume that the particles are located at height $h$ measured from the lower SiO$_2$ interface. For a given gap distance, $d$, and assuming a particle radius $a = 30\,$nm, the distances of the particle surface from the lower and upper confining surface is $x_L = d-h-a$ and $x_U = h-a$, respectively.

For the particle modeled by a sphere, the effective potential $\psi_{P,eff}$ is obtained from \cite{adamczyk1996role}
\begin{equation}
\label{Sphereiso}
\psi_{P,eff} \frac{q_e}{k_B T} = 8*\tanh(\overline{\psi_P}/4) \left(1+\sqrt{1-\frac{2 A+1}{(A+1)^2} \tanh(\overline{\psi_S}/4)^2}\right)^{-1},
\end{equation}
using the dimensionless potential $\overline{\psi_P} = \psi_P q_e/k_B T$ and a dimensionless particle radius $A = a \kappa$. In Eq. (\ref{Sphereiso}), $q_e$ is the elementary charge and $k_B T$ is the thermal energy. Similarly, for the isolated surface the effective potential $\psi_{S,eff}$ is obtained from the dimensionless plane potential $\overline{\psi_S}$ by
\begin{equation}
\label{PsiSurface}
\psi_{S,eff} \frac{q_e}{k_B T} = 4 \tanh(\overline{\psi_S}/4).
\end{equation}
The interaction energy $\Phi(h)$ between a sphere and two similar surfaces with separations $x_L = d-h-a$ and $x_U = h-a$ is finally obtained from
\begin{equation}
\label{phitotal}
\Phi_0 = 4 \pi \epsilon \epsilon_0 a \psi_{P,eff} \psi_{S,eff},
\end{equation}
and
\begin{equation}
\label{phivsh}
\Phi(h) = \Phi_0 \left(e^{-(d-h-a) \kappa} + e^{-(h-a) \kappa}\right).
\end{equation}
Eqs. (\ref{Sphereiso})-(\ref{phivsh}) describe the interaction energy of a sphere in nano-fluidic confinement between similar surfaces. At the timescales investigated here, the system is in thermodynamic equilibrium and therefore the interaction energy give direct access to the occupation probabilities $p_{sim}(h)$ of a particle residing at height $h$ via the Boltzmann distribution:
\begin{equation}
\label{probability}
p_{sim}(h) \propto e^{-\Phi(h)/k_B T}.
\end{equation}
To compare Eq. (\ref{probability}) with the experimental results, we obtain an experimental probability $p_{exp} (h)$ from the $h(d)$ data shown in Fig. \ref{fig:height}c by calculating normalized occupation histograms for $d$ and $h$ intervals of $4\,$nm. Because of the periodic nature of the particle contrast, we often cannot assign single-valued heights to a particular particle contrast. Instead, when we calculate $p_{exp}$, we weigh all possible heights for a given contrast as a single measurement. As a result the "probability" $p_{exp}(h)$ shown in the left column in Fig. \ref{fig:LSA}b does not add up to unity for a given gap distance, $d$, but rather to the average number of possible $h$-solutions for each contrast. However, the procedure ensures that the probability for the physical branch of $h$-values can be compared with the model calculations shown in the right column in Fig. \ref{fig:LSA}b titled "$p_{sim}(h)$". For fitting the model to the data we generate a simulated map of particle height "probabilities" including the unphysical height positions for a given particle contrast. For this we simply mirror the simulated particle probabilities $p_{sim}(h)$ at the experimentally determined height values of maximum particle contrast. The result of this procedure is shown in Fig. \ref{fig:LSA}b in the middle column.

The fits were performed using a non-linear least square fit method in the range of gap distances from 240 to 400$\,$nm. The range was restricted to minimize errors from the very narrow distributions for $d < 240\,$nm, where the width of the distributions is comparable to the diffusion length during illumination. The surface potential of the silicon oxide interfaces was approximated from literature \cite{Behrens2001} to be $\psi_S = -67\,$mV. We resort to a two step fitting procedure to enhance the convergence of the fits. In a first step we fixed the surface potential $\psi_P$ for all particles to the experimentally measured zeta potential $\psi_\zeta = -58\,$mV. Under this condition we obtained a Debye length of $\kappa^{-1} = 23.4 \pm 0.1\,$nm from a global fit to all particles. In a second step we fixed $\kappa$ and used the individual particle potentials as fit parameters. The obtained values of $\psi_P = -106 \pm 8, -54 \pm 4,$ and $-65 \pm 5\,$meV and the corresponding simulation data are shown in Fig. \ref{fig:LSA}b and c from top to bottom.

Average cross-sections through $p_{sim}(h)$ and $p_{exp}(h)$ for a gap distance range of $350-400\,$nm are show in Fig. \ref{fig:LSA}c. The simulated result is shown as red line. The mirrored and the summed signals are shown in yellow and black, respectively. There is good agreement of fit and data for a distance $h(d)-h_{mirr} > 20\,$nm, as expected from the diffusion length within the illumination interval of $\approx 40\,$nm.

The quality of the agreement with the LSA approximation further validates our optical model. The quantities obtained for the physical parameters are in accordance with the expected values, and the surface potentials for the individual particles could be obtained.

\section{Conclusions}

%For confinement distances between 130 and 270 nm the particle height is measured with a precision below 3 nm. At turn around points of the contrast the precision is $\approx 10$ nm and due to the periodicity of the interference signal,  multiple height solutions exist for a particular contrast. In case the height distribution is more confined, some of these solutions can be excluded because they are physically unlikely. The high speed measurement (2 ms acquisition time per frame and 100 $\mu$s effective exposure time per particle) allows us to quantify the confinement energy landscape and deduce a value for the Debye length, $\kappa^{-1}$, of the system and also estimate the surface potential.

We have established a parametric model to describe the contrast measured by iSCAT imaging of individual particles in a tunable nano-fluidic gap. The model describes the gap-height dependent signals detected from particles immobilized at the confining surfaces and obtained from individual diffusing particles with a consistent set of parameters. For the freely diffusing particles, the parameters are extracted from a measurement of the particle and background signal as a function of the nano-fluidic gap distance. Two of the parameters are obtained from the envelope of the intensities recorded. The third parameter, i.e., the particle scattering phase, $\phi_0$, is obtained at small separations when the confinement of the particle is restricted to a height range of just a few nanometers.

As a result, the particle contrast is calibrated \textit{in-situ} and the model parameters are obtained without further calibrations. In a range of gap distances the parameters can directly be mapped to a unique particle height via the model. Because of the periodic particle contrast as a function of particle height multiple solutions exist at larger gap distances of $d > 250\,$nm. The accuracy of the height measurement is given by the accuracy of the gap-distance measurement, the errors in the parameter determination, and by the intensity fluctuations. We estimate an accuracy for the gap-distance measurement of $\lesssim 2\,$nm from the accuracy of the z-calibration and the laser noise. The error arising from the uncertainty in $\phi_0$ can be estimated from the statistics of the confined particle distribution to $\lesssim 4\,$nm. The errors due to the parameters $\gamma$ and $p$, and from the intensity fluctuations can be estimated from the precision of matching the model envelope to the recorded data to a statistical error of $\approx 2\,$nm and a systematic error of about $\approx 2\,$nm at intermediate particle contrast values. Thus we estimate that absolute heights can be measured in the single valued contrast regions with an accuracy of $\approx 10\,$nm and a precision of $\approx 2\,$nm at intermediate contrast values. Naturally, the error increases significantly at the extremal values of the contrast signal.

We apply the measured particle height distribution to obtain the physical parameters of the system describing the interactions between particle and surfaces. For this we calculate the height probability distribution for the individual particles, which are directly connected to the free energy of the system. Here the short illumination time of $\sigma_{illu} \approx 100\,\mu$s achieved by a single-pass laser scan per frame is critical to arrive at meaningful height values. Thus, the NCA allows us to observe the narrowing of the particle height distribution at increasing confinement. Consequently, the physical parameters of the system are obtained, i.e., the Debye length, and the surface potentials. Their values are in agreement with the values obtained from measurements of the mean zeta potential and literature values of the SiO$_2$ surface potential.

We emphasize that the experiment employing the NCA grants \textit{in-situ} access to the free energy landscape of the system in all three dimensions at the individual particle level and as a function of confinement. This opens up new possibilities to study in great detail the behavior of individual confined nano-objects as a function of system parameters. These parameters include not only the nature of the particles and confining surfaces, but also the application of external electric or magnetic fields. Similarly, laterally inhomogeneous surfaces patterned by topography, charge or magnetic moments are also accessible. Using the NCA, these systems can be studied as a function of confinement, which will be important to separate effects arising from the charged surfaces and the electic/magnetic fields.

\section*{Acknowledgments}
The authors thank U. Drechsler, M. Tschudy, and S. Reidt for technical support, C. Bolliger for proof-reading, and U. Duerig and M. Krishnan (University of Zurich) for fruitful discussions. Funding has been provided by the European Research Council StG no. 307079.

\renewcommand{\thefigure}{A\arabic{figure}}
\section*{Appendix A: Calibration of z-stage}
\setcounter{figure}{0}

In this section we describe the calibration of the piezo fine-positioning stage holding the cover-glass. The calibration is done by removing the objective, lenses and AOD from the setup (Fig. \ref{fig:setup}a). The collimated laser beam is reflected at the glass-air and air-silicon interfaces. The resulting interference signal is measured by the camera.
A calibration factor of $z_{\mathrm{cal}} = 0.984 \pm 0.001$ is obtained from the distance of interference extrema (the error is given by the standard error of the measurement). This value is confirmed by a second measurement using a commercial interferometer (IDS3010, Attocube) on a silicon interface mounted on the piezo stage.

\renewcommand{\thefigure}{B\arabic{figure}}
\section*{Appendix B: Interference of a focused laser beam by the glass-air-silicon slit}
\setcounter{figure}{0}
In our illumination scheme using a scanned focused laser beam the incident light rays depart from normal incidence. As a result the periodicity of the observed intensity modulations as a function of glass-silicon distance is greater than expected from a simple normal incident model. In order to tackle the problem we calculate in this section the interference pattern observed using a high NA objective and the fit to measured data. This "calibration" is done in air, to avoid pressure effects from the contained water volume in the filled system.

\begin{figure} \centering
	\includegraphics{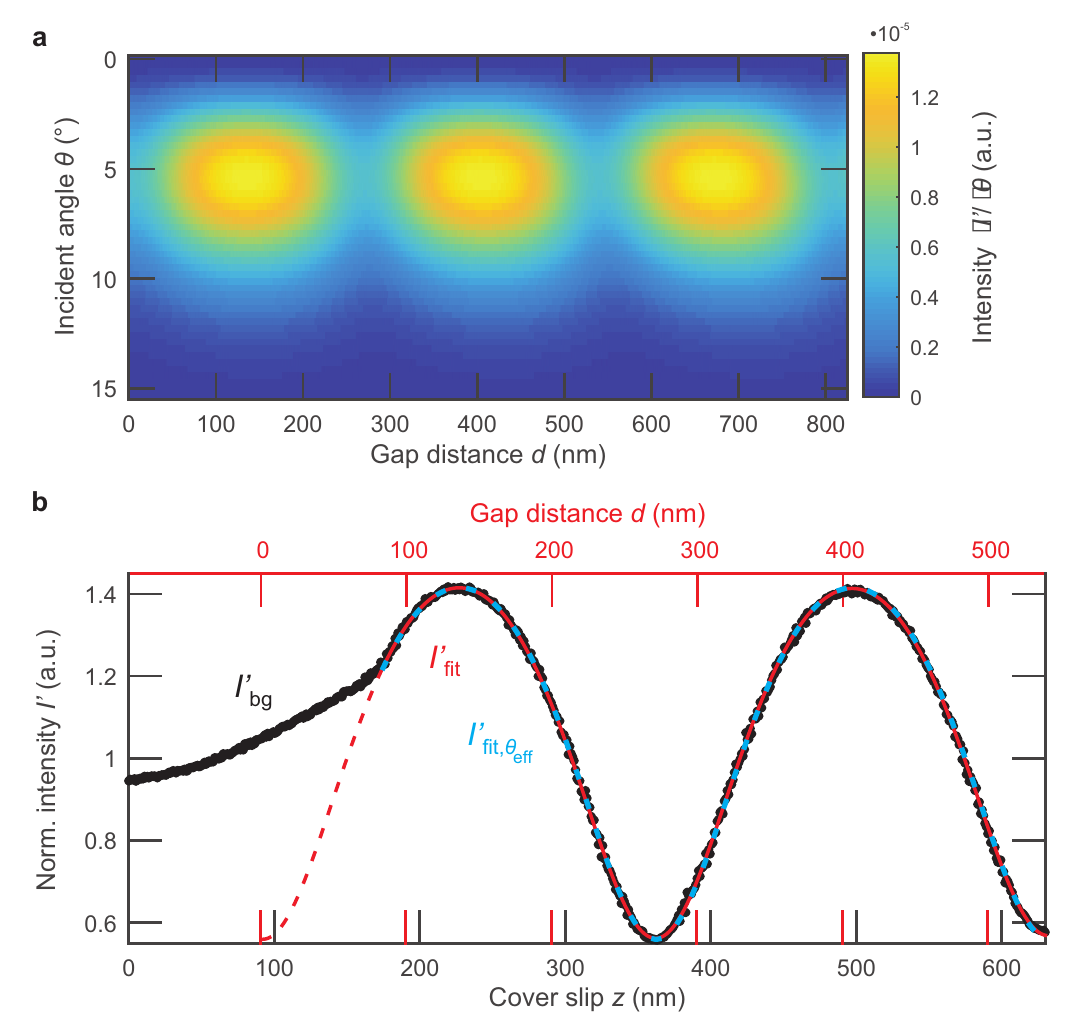}
	\protect\caption{The effect of the incident angle due to focusing of a Gaussian beam on the measured intensity. \textbf{a}) Simulated intensity for varying gap distance of a glass-air-silicon nanoslit. The exponential decay for larger incident angles is caused by underfilling the objective with a Gaussian beam, while the integration of the azimuthal angle is weighting the intensity towards higher angles. \textbf{b}) Measured signal from the background of a glass-air-silicon nanoslit (black) while moving the cover-glass along optical axis. The dashed red line indicates a fit as a function of gap distance $d$ considering all incident angles. The dashed blue line is simulated with the same parameters but considering only the effective angle $\theta_{eff}$.  \label{fig:eff_div_angle}}
\end{figure}

We describe the laser beam by a paraxial Gaussian beam centrally incident on the back aperture of the objective.
We follow the theory by Richards \cite{Richards1959} describing the reflection of a focused optical field by a high NA objective. After reflection and refraction by the aplanatic optical lens the electric field distribution in the collimated farfield is given by
\cite{novotny2006principles},	

\begin{equation}
\label{reffield}
E = -\pi E_0 e^{2ik\cos\left(\theta\right) z_0} \int_{0}^{\rho_{max}}e^{-\rho^2/w_0^2}\rho\left[ r_p\left(\theta\right) -  r_s\left(\theta\right)\right]d\rho.
\end{equation}
Where $\rho$ is the distance from the optical axis in cylindrical coordinates and $z_0$ is the distance between focus and interface. The Fresnel reflection coefficients $r_p\left(\theta\right)$ and $r_s\left(\theta\right)$ are functions of the incident angle $\theta$ in the focused beam. The mapping to $\rho$ in the collimated beam is given by the sine law $\sin\left(\theta\right) = \rho / f$, where f is the focal distance of the objective.
In order to arrive at the reflected field in the case of our glass-air-silicon system we use the transfer-matrix-method to calculate the reflection
coefficient $r_p\left(\theta\right)$ and $r_s\left(\theta\right)$. The light intensity distribution is shown in Fig. \ref{fig:eff_div_angle}a as a function of gap distance, $d$, and $\theta$. As expected the gap distance modulates the overall light field. In angular direction we observe a maximum intensity at a finite divergence angle $\theta$ which is approximately independent of the gap distance, $d$.

We approximate the signal observed by the camera due to the scanning laser spot by the integrated intensity contained in the collimated beam. An excellent fit (see Fig.\ref{fig:eff_div_angle}b) to the measured data can be obtained using just two fit parameters, the Gaussian-beam width, $w_0$, and the gap distance zero position $d_0$. For the refractive indices of the stack, $n_{\mathrm{glass}} = 1.52$, $n_{\mathrm{air}} = 1.00$, $n_{Si} = 4.14$ obtained from literature were used. We obtain a beam width of 0.42 mm, giving rise to the strong underfilling of the objective and the fast decay of light intensities at increasing incident angles shown in Fig. \ref{fig:eff_div_angle}a.

The narrow angular distribution and the gap distance independence allows us to introduce an effective incident angle $\theta_{eff} = 5.9 \pm 0.1 ^\circ$ to describe the system \cite{Pliskin1968}. This reduces the description of the optical response of the system to one incident angle and thus speeds up the numerical calculation considerably. We can use the effective angle to calculate the observed periodicity of the interference pattern measured by the camera. We obtain an effective increase of the observed periodicity by $0.68 \%$ in water as compared to normal incidence.

The model and the excellent fit to the data in air allow us to predict the light intensities observed in the water filled system. Because of pressure effects from compressing the droplet confined between glass and silicon sample, the glass-silicon distance depends on the history of the experiment. Therefore we obtain the actual gap distance by comparing the normalized background intensity to the intensity calculated from the effective incident angle model, see Fig.\ref{fig:setup}d of the manuscript.

\renewcommand{\thefigure}{C\arabic{figure}}
\section*{Appendix C: Image processing}
\setcounter{figure}{0}
In this appendix we describes how the raw images are normalized and how the background intensity at the particle positions are derived. First, a dark image is measured at zero laser power and subsequently subtracted from each frame of the entire raw image stack.
We estimate the background intensities at the particle positions (see blue data in Fig. \ref{fig:setup}d) from averaging the pixel intensities in the spatial and temporal neighborhood. The intensity values at pixel positions corresponding to tracked particle positions are excluded from the average.

In the case of immobilized particles the average is taken over the spatial neighborhood by applying a median average filter with a filter window of 30 x 30 pixels, much bigger than the observed particle size of 7 x 7 pixels. In the case of moving particles the particles reside at a particular location for just a few frames because of lateral diffusion. Therefore the temporal frame position and peak intensities of the background interference extrema can still be derived from least-squares second-order polynomial fits of the remaining intensity values versus frame time. We match the temporal frame position and the normalized peak intensities to the calculated intensities for the glass-water-silicon plan parallel system (Eq. \ref{Ibg}). As a result we obtain the conversion of frame time to gap distance $d$. The desired background intensity at the particle position is then obtained by interpolation at the respective gap distance $d$.

\renewcommand{\thefigure}{D\arabic{figure}}
\section*{Appendix D: Effect of focus position}
\setcounter{figure}{0}
\begin{figure} \centering
	\includegraphics{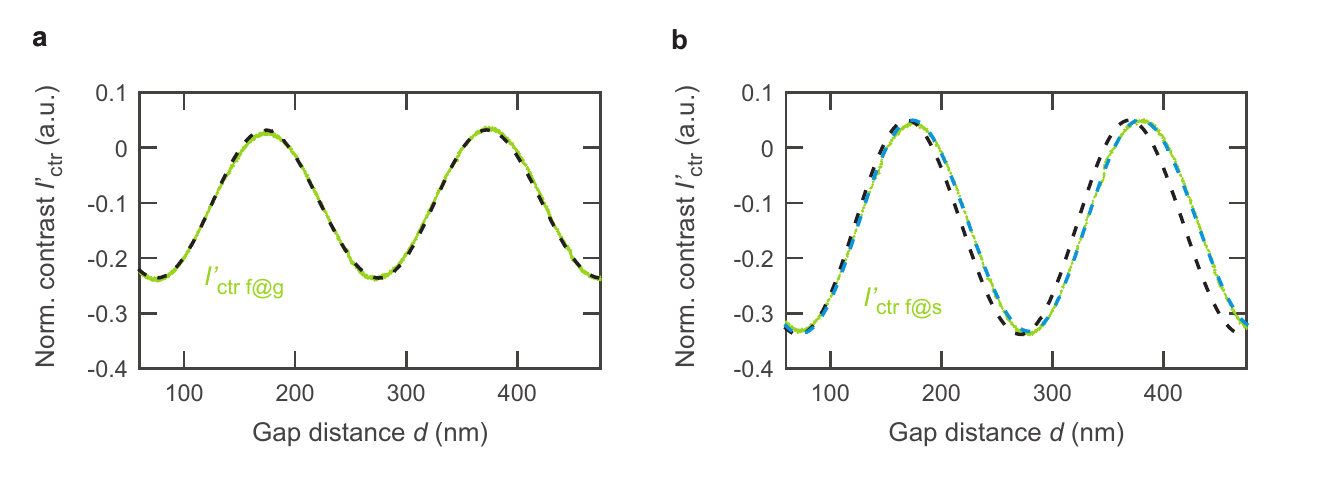}
	\protect\caption{The effect of the focus position on the particle contrast. \textbf{a,b}) Examples of measured contrast signals for a particles attached to the cover glass (green). The dashed black lines represent fits according to our optical model without considering defocussing. \textbf{a}) The height of the objective is adjusted so that the particle at the cover glass stays in focus. \textbf{b}) The height of the objective is not changed so that a particle at the substrate stays in focus. The dashed blue line corresponds to a fit in which an additional linearly increasing phase shift is considered.\label{fig:defocusing}}
\end{figure}

Two experiments with particles fixed to the cover-glass are carried out to investigate the effect of the focus position. In the first measurement, the height of the cover-glass and the objective are moved in parallel with a fixed distance, and thus the sticking particles stay in focus as the gap distance changes. The contrast behavior of these particles (Fig. \ref{fig:defocusing}a, green data) can be described by our model (dashed black line) without correction. In the second measurement, a particle sticking to the sample surface is in focus and the height of the objective is constant, while varying the gap distance. The contrast oscillates for these particles (Fig. \ref{fig:defocusing}b, green data) with a longer wavelength than in our model (dashed black line). This effect may originate from the Gouy phase and is phenomenologically addressed by adding a phase shift to the scattering phase in Eq. (\ref{rpg}) that linearly increases with the distance of the particle from the focus plane at the substrate $\Delta \phi =  z_f \pi (h-a)n_{H_2O}/\lambda$. At a factor of $z_f=0.117$, the simulated contrast (dashed blue line) coincides with the measured data. This effect has also been considered for the suspended particles.

\newpage
\renewcommand{\thefigure}{E\arabic{figure}}
\section*{Supplementary materials}
\setcounter{figure}{0}
$\textcolor{blue}{Movie\,1\label{vis1}}$\\
\begin{figure}[h] \centering
	\includegraphics{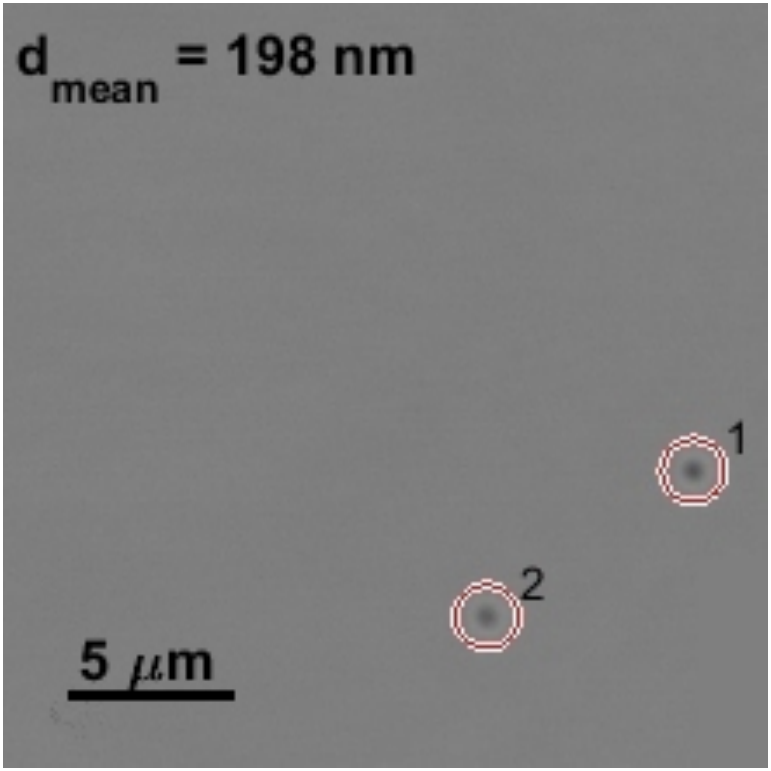}
	\protect\caption{The video shows exemplarily the contrast of two suspended $60\,$nm gold spheres. The average gap distance, $d_{mean}$, between the two confining surfaces is increased during the video. The red rings visualize the central particle position obtained by radial symmetry-based tracking \cite{Parthasarathy12natmet}. The playback speed is reduced by a factor of 2 compared with the real-time acquisition.\label{fig:movie1}}
\end{figure}

\end{document}